\documentclass[
aps,pre, 
twocolumn,
reprint,
amsmath,amssymb,
superscriptaddress,
notitlepage
]
{revtex4-2}

\usepackage{graphicx}
\usepackage{dcolumn}
\usepackage{bm}
\usepackage{graphicx}
\usepackage[dvipsnames]{xcolor}
\usepackage{bm}
\usepackage{mathtools}

\usepackage[american]{babel} 
\usepackage[T1]{fontenc}             
\usepackage{lmodern}                 
\usepackage[utf8]{inputenc}          
\usepackage[babel]{csquotes}         
\usepackage{graphicx}                
\usepackage{pgfplots}
\usepackage{booktabs}               
\usepackage{listingsutf8}         
\usepackage{adjustbox}
	
\usepackage{pgf}

\usepackage{graphicx,xcolor}
\usepackage{tabularx}

\newcommand{\eps}{\epsilon}

\begin{document}

\preprint{AIP/123-QED}

\title[Equivalence of Additive and Multiplicative Coupling in Spiking Neural Networks
]{Equivalence of Additive and Multiplicative Coupling in Spiking Neural Networks}

\author{Georg Börner}
 \affiliation{Chair for Network Dynamics, Institute for Theoretical Physics and Center for Advancing Electronics Dresden (cfaed), TU Dresden} 
 \author{Fabio Schittler Neves}
 \affiliation{Chair for Network Dynamics, Institute for Theoretical Physics and Center for Advancing Electronics Dresden (cfaed), TU Dresden} 
\author{Marc Timme}
 \affiliation{Chair for Network Dynamics, Institute for Theoretical Physics and Center for Advancing Electronics Dresden (cfaed), TU Dresden} 
\date{\today}
\begin{abstract}
Spiking neural network models characterize the emergent collective dynamics of circuits of biological neurons and help engineer neuro-inspired solutions across fields. Most dynamical systems' models of spiking neural networks typically exhibit one of two major types of interactions:  First, the response of a neuron's state variable
to incoming pulse signals (spikes) may be additive and independent of its current state. Second, the response may depend on the current neuron's state and multiply a function of the state variable.
Here we reveal that spiking neural network models with additive coupling are equivalent to models with multiplicative coupling for simultaneously modified intrinsic neuron time evolution.
As a consequence, the same collective dynamics can be attained by state-dependent multiplicative and constant (state-independent) additive coupling. 
Such a mapping enables the transfer of theoretical insights between spiking neural network models with different types of interaction mechanisms as well as simpler and more effective engineering applications.

 \end{abstract}

\maketitle

\section{Background}
Differential equations model the time evolution of a broad range of natural and human-made systems with time-continuous intrinsic dynamics and interactions.
\cite{Birkhoff1927DynamicalSystems, Strogatz2018NonlinearDynamicsChaos, 
Hansel1995SynchronyExcitatoryNetworks,
Ashwin2004EncodingConjugateSymmetries}.
The dynamics of systems with short-lasting interactions may be modeled by networks of artificial pulse-coupled or spiking neurons. They constitute hybrid dynamical systems where 
the intrinsic continuous time-evolution of the units' dynamics are interrupted by time-discrete events. At these events, the incoming pulses (spikes) deterministically change the state variables of the units in the receiving end. The event times, in turn, are determined by the state space trajectory passing certain subsets, e.g., crossing some manifold or hitting boundary points of state space.
Since half a century, spiking and pulse-coupled network models play key roles in our understanding of natural phenomena such as the emergence of waves or synchrony \cite{Mirollo1990Synchronization, Maass1997SpikingNetworks, Wang2010ReviewPulseCoupled}. They also help to design appropriate collective dynamics for engineered systems implementing desired functionalities such as distributed sensing \cite{Yanmaz2018DroneNetworks, Klinglmayr2012GuaranteeingGlobalSynchronization, Klinglmayr2016ConvergenceSynchronization}.

Due to their hybrid nature, spiking systems exhibit, and thus capture, novel phenomena not present in systems of smooth, time-continuous differential equations. Examples include the emergence of linearly unstable attractors (in the sense of Milnor) \cite{Milnor1985Attractor, Timme2002PrevalenceUnstableAttractors,Ashwin2005UnstableAttractorsExistenceRobustness}, the existence of speed limits in the relaxation dynamics of networks of spiking units \cite{Timme2004TopologicalSpeedLimits, Van2023TopologicalSpeedLimit}, the possibility of identical oscillators overtaking each other even though they are symmetrically coupled \cite{Kielblock2011BreakdownOrderPreservation}, and the emergence of isochronous regions where multiple periodic orbits of the same period coexist 
\cite{Li2017IsochronousDynamicsDelay}.

These phenomena have been demonstrated for models where the interactions are additive and the coupling strength is constant, i.e. independent of the current state of the unit that receives an interaction pulse signal. 

However, several models of biological neural circuits exhibit state-dependent, multiplicative coupling
\cite{Hodgkin1952QuantitativeMembraneCurrent, Winfree1967BiologicalRhythmsPopulations,
Ermentrout1996TypeIMeurons,
Brette2007SimulationNetworksSpikingNeurons}
and recent conceptual works on artificial computational devices demonstrated that networks of pulse-coupled oscillatory neurons with multiplicative coupling give rise to novel computational features such as reconfigurability and improved robustness, see, e.g.,  \cite{Neves2020ReconfigurableComputation} and \cite{Wang2022MemristiveLIFNeuron}.
In addition, certain types of multiplicative coupling may be simpler and more efficient to implement in hardware.

Here we demonstrate that under certain conditions, network models with state-dependent, multiplicative coupling and models with constant additive coupling are equivalent in the sense that they exhibit identical spiking dynamics. The implications of this result are twofold. First, many theoretical and practical results on oscillator networks with additive coupling can be transferred to those with multiplicative coupling and vice versa, by transforming the neuron model accordingly.
In particular, the extensive body of work on additive coupling systems can be carried over effectively to multiplicative coupling systems, while phenomena that have been described for systems with multiplicative coupling, such as  robust and reconfigurable computation over a combinatorial number of inputs  observed in \cite{Neves2020ReconfigurableComputation} should also be expected for classes of additive-coupling systems.
Second, for technical applications such as the growing field of analogue and spiking computation, depending on the particular neuron model and the coupling network, usually one type of coupling can be designed and implemented more effectively. For example, systems with conductance-based leaky integrate-and-fire (IF) neurons allow for straightforward multiplicative coupling.
Using the connections established in this paper, one can choose the coupling type more freely by modifying the neuron model accordingly.

\section{Pulse-coupled Oscillator Networks and Phase Formalism}

Consider an $N$-dimensional dynamical system with states $x_i(t)$, $i \in \{1, \dots, N\}$,  with a dynamic  defined by $N$ coupled ordinary differential equations  
\begin{equation}\label{eq:1}
\frac{dx_i}{dt} = f_i(x_i) + S_i(\{x_j\},t)
\end{equation}
with continuous functions $f_i(x_i): \mathbb{R} \to \mathbb{R}$, and interaction mediated by the terms $S_i(\{x_j\},t)$,
where $\{x_j\}=\{x_1, \dots, x_N \}$.
    In the following, we consider oscillatory neurons as one core example
    \cite{Mirollo1990Synchronization, Ernst1995DelayInducedSynchronization, Ernst1998DelayInducedMultistableSynchronization, Memmesheimer2006DesigningDynamicsNeuralNetworks,Abbott1993AsynchronousStates,Memmesheimer2010StableUnstableOrbitsDelays, Timme2002PrevalenceUnstableAttractors, Timme2008SynchronyStable},  where variables $x_i$ are typically interpreted as a potential.
    For implementing periodic free dynamics we require that  $f_i(x_i): \mathbb{R} \to \mathbb{R}^{+}$ is positive and state a reset mechanism 
    \begin{equation}
    \label{eq:reset_x}
x_{i}(t_{i,m}^{+}):=x^{\mathrm{reset}}\equiv 0,
\end{equation}
which sets the state variable  $x_i$ to a reset value $x^{\mathrm{reset}} \coloneqq 0 $
at the discrete times $t_{i,m}$ where  it passes a threshold value $x^{\mathrm{thr}}_i\equiv1$ for the $m$-th time (from below),
\begin{equation}
\label{eq:threshold_x}
x_{i}(t_{i,m}) = x^{\mathrm{thr}}\equiv 1\, ,\, \, \, \left. \frac{dx_{i}(t)}{dt}\right| _{t=t_{i,m}}>0.
\end{equation}
 Together with the intrinsic dynamics defined by equation \eqref{eq:1} for $S_i=0$, the reset mechanism
 \eqref{eq:reset_x} creates a free period $T_i$ for each neuron.
  Note that the choice of
\( x^{\mathrm{thr}}\equiv 1 \) and \( x^{\mathrm{reset}}\equiv  0 \) is made without
loss of generality.  
    The interaction between different neurons, as described by the functions $S_i(\{x_j\},t)$, can take on many forms. Short-lasting, momentary interactions as found in biological systems such as populations
of flashing fireflies or networks of spiking neurons in the brain, are often  adequately captured in terms of pulse coupling.
  Within this context,
    a neuron that reaches its threshold is said to \enquote{fire} or \enquote{spike}, and sends a stereotyped short-lasting signal,
    which mediates its effect on connected units.
  Such 
pulse coupling between oscillatory neurons can generally be written as \begin{equation}
\label{eq:internal_current}
S_{i}(t)= - \sum _{\substack{j=1 \\ j\neq i}}^{N}\sum _{m=-\infty }^{\infty }\eps_{ij}K_{ij}(t-t_{j,m}-\tau_{ij})
\end{equation}
 where \( \eps _{ij}\in \mathbb{R} \) quantifies the strength of the coupling from neuron
\( j \) to neuron \( i \) and the response kernels \( K_{ij}(t) \)
satisfy \( K_{ij}(t)\geq 0 \), \( K_{ij}(t)=0 \)
for \( t<0 \), and \hbox{\( \int _{-\infty }^{\infty }K_{ij}(t)dt=1 \)}. 
\( t_{j,m} \) is the $m$-th time  at which unit \( j \)
reaches the threshold \( x^{\mathrm{reset}}=1 \), see equation \eqref{eq:threshold_x} and $\tau_{ij}\geq 0$ is the time it takes for a pulse to travel from neuron $j$ to neuron $i$.  Note that for $\tau_{ij}>0$ additional rules regarding the order of processing spiking events may be necessary.
Often, pulse-coupled systems exhibit a strong time scale separation between the interaction duration and the intrinsic time scales of the neurons  so that the kernels \( K_{ij} \) in (\ref{eq:internal_current}) may
be idealized as a  Dirac distribution \begin{equation}
\label{eq:K_delta}
K_{ij}(t)=\delta (t)
\end{equation}
 such that the coupling \eqref{eq:internal_current} becomes discontinuous.
In this work, we focus on purely inhibitory coupling, i.e., $\epsilon_{ij}\geq 0 $. 
For simplicity of presentation, in the following we set
 $\tau_{ij} =  0$ (instantaneous interactions). However, we point out that our findings   easily generalize  to   delayed systems.

\begin{figure}[!]
    \centering
-    \includegraphics[scale=1]{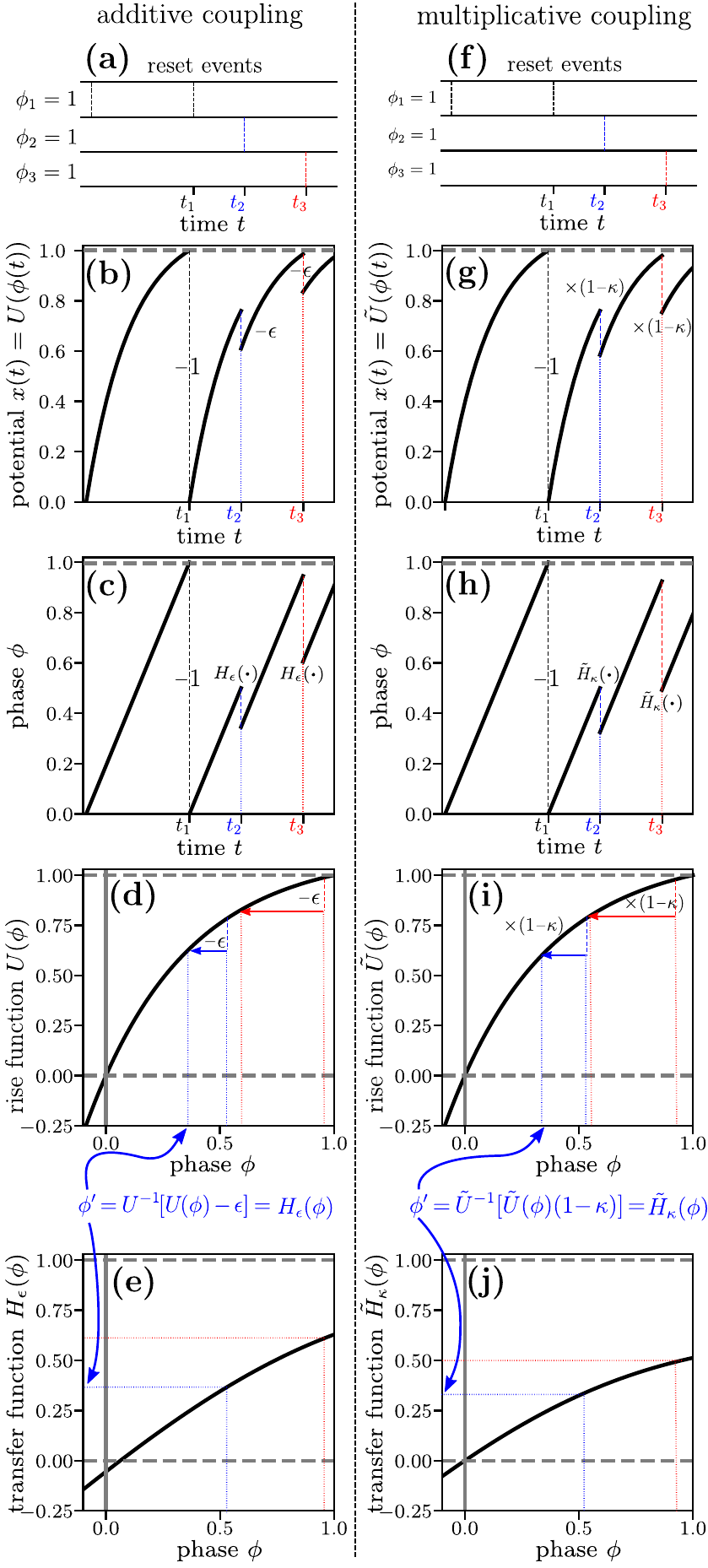}
    \caption{
    \textbf{Phase formalism for (a-e) additive and (f-j) multiplicative inhibitory pulse coupling.}
    (a,f) Reset event patterns of three-unit network as function of time.
     (b,g) Rise function of first neuron  $x_1(t) \equiv x(t) $. At $t_1$, there is a threshold-induced reset, and at $t_2$ and $t_3$ resets of other neurons cause a discontinuous jump in $x(t)$.
     (c,h) The same dynamics represented in terms of the phase $\phi_1(t) \equiv \phi(t)$.
     It evolves with constant velocity $d\omega_1/dt = 1/T_1$, interrupted by discontinuous reset and interaction events.
     (d,i) Rise functions $U(\phi)$, $\tilde{U}(\phi)$ as functions of the phase $\phi$. The phase jumps in blue and red are mediated in terms of jumps in the rise functions  $U(\phi)$, $\tilde{U}(\phi)$, respectively.
     (e,j) Transfer functions $H_\epsilon (\phi)$, $\tilde{H}_\kappa (\phi)$ summarize the effect of an incoming pulse on the phase $\phi(t)$ of the receiving neuron.
    }
\label{fig:MS_approach}

\end{figure}

The effect of neuron $j$ reaching threshold at time $t_j$ (dropping the index $m$ for sake of readability)  can be summarized as instantaneous, discontinuous change of the other neurons' states $x_i$, $i \neq j$ by a constant $\epsilon_{ij}$, which is independent of both the time and the state $x_i$ itself:
\begin{equation}\label{eq:update_rule}
 x_i(t_j^+) = 
 x_i(t_j) -  \epsilon_{ij},
\end{equation}
for illustration see figure \ref{fig:MS_approach}a,b.
Note that large coupling strengths $\epsilon_{ij}$ may lead to a negative potential $x_i(t_j) -  \epsilon$ just after a reset, requiring equation  \eqref{eq:1} to provide solutions with $x_i<0$ for the non-interaction case $S=0$.
This may pose a modeling problem, as naturally occurring neurons typically have a finite lower boundary regarding their potential \cite{Timme2008SynchronyStable,Timme2002CoexistenceRegularIrregularDynamics}. 

  The resulting dynamical system is of hybrid type with continuous-time free (uncoupled) time evolution of neuron states interrupted by one of two types of events occurring at discrete times, where maps are applied: i) reset (eq. \eqref{eq:reset_x}) or ii) spike reception (eq. \eqref{eq:update_rule}).
The hybrid nature enables to transfer to new neuron state variables that are time-like (phases), in terms of the phase formalism introduced by Mirollo-Strogatz \cite{Mirollo1990Synchronization}. Essentially a nonlinear transformation, it  parametrizes the free time evolution of the oscillatory neurons in terms of periodic phases $\phi_i(t)$ that evolve with a constant phase velocity $d\phi_i/dt = \omega_i$ until they reach the threshold value $\phi_i^{\mathrm{thr}} \coloneqq 1$ and are reset to $0$. 
The interaction between different neurons is mediated via a so-called {rise function} or  {neuron potential}  $U_i(\phi_i):\, \phi_i \to (-\infty, 1]$, which is monotonically increasing, twice continuously differentiable and typically normalized to $U_i(0)=0$ and $U_i(1) = 1$.
Then, the response of unit $i$ on another unit $j$ reaching the threshold value $\phi_j = 1$ at time $t_j$ is defined as
\begin{equation}\label{eq:update_rule_MS}
\phi_i (t_j^+) \coloneqq
 H_{\epsilon ,i}(\phi_i(t_j)),
\end{equation}
with  \textit{transfer function} 
 \begin{equation}\label{eq:transfer_function}
H_{\epsilon,i}(\phi_i) \coloneqq U_i^{-1}[U_i(\phi_i)-\epsilon],
\end{equation}
with $\epsilon \equiv \epsilon_{ij}$.
 For a free dynamic of the form ${dx_i}/{dt} = f_i(x_i) $, as in equation \eqref{eq:1}, it is always possible to choose  
 \begin{equation}\label{eq:potential_function}
 U_i(\phi_i) \equiv x_i( \phi_i T_i),
 \end{equation} where $x_i(t)$ the free solution 
  with period $T_i$.
 The phase description of the free time evolution is then given by
 \begin{equation}
  \frac{d\phi_i(t)}{dt} = {\omega_i} = \frac{1}{T_i}, 
  \qquad \phi_i(t + T_i) = \phi_i(t),
 \end{equation}
 where $\phi_i(t)$ is reset to $0$ after reaching $1$,
$
 \phi_i(t_i^+) \coloneqq 0    
$.
 For illustration, see figure \ref{fig:MS_approach}a-e.

As a standard example, consider leaky integrate-and-fire neurons as described by 
 \begin{equation}\label{eq:free_dyn_IF}
\frac{dx}{dt} 
= I - \gamma x,
\end{equation}
(dropping the index $i$ throughout, as we focus on one specific neuron at a time here)
with  positive constants $I$ and $\gamma<I$,
giving the free time evolution
\begin{equation}
x(t) = \frac{I}{\gamma}(1-e^{-\gamma t}) \qquad \mathrm{for} \qquad 0<t\leq T
\end{equation} and
 $x(t+nT)= x(t)$ for $n \in \mathbb{Z}$,
with free period length $T = - \gamma^{-1} \textrm{ln}(1-{\gamma}/{I})$.
According to equations \eqref{eq:potential_function} and  \eqref{eq:transfer_function}
  one finds the rise function
 \begin{equation}\label{eq:IF_U}
 U^{\mathrm{IF}}(\phi) = \frac{I}{\gamma}\left(1- e^{-\gamma\phi T}\right)  =  \frac{I }{\gamma}\left(1- \left(1-\frac{\gamma}{I }\right)^{\phi}\right),
 \end{equation}
  and the transfer function
   \begin{equation}\label{eq:IF_H}
H_{\epsilon}^{\mathrm{IF}}(\phi)   = \frac{\ln\left((1-\gamma/I)^{\phi} + \epsilon \gamma/I\right)}{\ln(1-\gamma/I)}.
 \end{equation}
Another common model class has been proposed by Mirollo and Strogatz \cite{Mirollo1990Synchronization} directly in terms of a rise function
\begin{equation}\label{eq:MS_U}
U^{\mathrm{MS, }b}(\phi) \equiv \frac{1}{b}\mathrm{ln}(1 + (e^b-1)\phi), \qquad  b>0.
\end{equation}
It is chosen such that, while $U^{\mathrm{MS, }b}$ itself is nonlinear and allows for realistic features such as down-concavity, the resulting transfer function is affine,
\begin{equation}\label{eq:MS_H}
H^{\mathrm{MS, }b}_{\epsilon}(\phi) = \alpha_\epsilon \phi + \beta_\epsilon,
\end{equation}
with constants $\alpha_\epsilon= e^{-b\epsilon}$ and $\beta_\epsilon = (e^{-b\epsilon}-1)/(e^b-1)$, 
 allowing for easy analytical treatment. 
 Note that 
 $U^{\mathrm{MS, }b}$ diverges already for a finite value $\phi_0 \equiv  1/(1-e^b)$: $\lim_{\phi \to \phi_0^+}   U^{\mathrm{MS, }b}(\phi)  = -\infty$, whereas  
 $\lim_{\phi \to - \infty} U^\mathrm{IF} (\phi) = -\infty$ for the integrate-and-fire neuron.  Hence, the effect of incoming pulses vanishes already at finite phases $\phi_0$, and the permitted phases are bounded, $\phi \in (\phi_0, 1]$, instead of $\phi \in (-\infty, 1]$, as for leaky integrate-and-fire neurons.

We emphasize again that the coupling considered up to this point is additive in the original potential-like variables $x_i$.

 \section{Multiplicative pulse coupling}\label{sec:multiplicative_pulse_coupling}

Although in many situations  the additive pulse coupling, as reviewed, in the last section is a fitting description, other choices may capture underlying mechanisms in existing systems better or can be implemented more effectively when designing systems.
In the following, we consider a type of inhibitory
pulse coupling between oscillatory neurons, in which the effect of an incoming pulse from neuron $j$ at time $t_{j}+\tau_{ij}$ is linearly related to the state $\tilde{x}_i(t_j)$ of the receiving oscillator,
\begin{equation}
\label{eq:S_multiplicative}
S_{i}(\tilde{x}_i, t)=-\tilde{x}_i(t) \sum _{\substack{j=1 \\ j\neq i}}^{N}\sum _{m=-\infty }^{\infty }\kappa_{ij}K_{ij}(t-t_{j}),
\end{equation}
where $\kappa_{ij} \in (0,1)$. 
 Throughout we use a tilde to point out multiplicative coupling, as opposed to additive coupling.
Again setting $K(t) = \delta (t)$, and assuming   instantaneous interactions, $\tau_{ij}=0$, for  simplicity of presentation,
  this leads to an update rule 
\begin{equation}\label{eq:update_rule_prop}
\tilde{x}_i(t_j^+) =(1 - \kappa_{ij})\, \tilde{x}_i(t_j)  ,
\end{equation}
instead of the additive-coupling update rule \eqref{eq:update_rule}.

In the following we 
modify the Mirollo-Strogatz formalism to also describe inhibitory multiplicative coupling. 
Given the free neuron dynamics in terms of a rise function $\tilde{U}_i(\phi_i):\,  \phi_i \to (-\infty, 1]$, we define a
multiplicative-coupling transfer function 
\begin{equation}\label{eq:transfer_function_prop}
 {\tilde{H}}_{\kappa,i}(\phi_i) \coloneqq \tilde{U}_i^{-1}[\tilde{U}_i(\phi_i)  (1-\kappa)],
\end{equation}
so that the effect of a  neuron $j$ resetting at time $t_j$ on the phase of a connected neuron $i$ is expressed as
\begin{equation}\label{eq:MS_mult}
\phi_i (t_j^+) \coloneqq 
  \tilde{H}_{\kappa,i}(\phi_i(t_j)),
\end{equation} with $\kappa = \kappa_{ij}$,
analogous to equation \eqref{eq:update_rule_MS} for standard additive coupling. 
 Now, the free time evolution is again defined by  \hbox{$d\phi_i/dt = \omega_i$} and a reset to $\phi_i=0$ upon reaching  the threshold value $\phi_i^{\mathrm{thr}} =1$, $\phi_i(t_i^+)\coloneqq0$.
 As with additive coupling, for given free neuron dynamics $\tilde{x}_i(t)$, one can  set $\tilde{U}_i(\phi_i) \equiv \tilde{x}_i(  \phi_i T_i)$ and $d\phi/dt = 1/T_i$, with the free period length $T_i$.

For multiplicative inhibitory pulse coupling, only neuron potentials with $\tilde{U}_i(\phi_i)\geq 0$ are sensible, such that $\tilde{U}_i(\phi_i) (1-\kappa_{ij}) \geq0$, avoiding that the sign of the induced phase jump changes in a physically implausible way.
Other than that, the requirements on $\tilde{U}_i$ are the same as for additive coupling, that is, $\tilde{U}_i$ should be monotonically increasing and twice continuously differentiable.
Again, generalization to systems with delayed coupling or inhomogeneous coupling strengths are straightforward.

As an example, we again consider the common Integrate-and-Fire model
 \begin{equation}\label{eq:IF_U_mult}
\tilde{U}^{\mathrm{IF}}(\phi) = \frac{I}{\gamma}\left(1- e^{-\gamma\phi T}\right)  =  \frac{I }{\gamma}\left(1- \left(1-\frac{\gamma}{I }\right)^{\phi}\right),
 \end{equation}
(again dropping the neuron index $i$ throughout for sake of readability)
 and find a multiplicative-coupling transfer function
\begin{equation}\label{eq:IF_H_prop}
\tilde{H}_{\kappa}^{\mathrm{IF}}(\phi) =   \frac{\ln\left(  
\kappa + (1-\kappa)    (1-\gamma/I)^\phi 
\right)}{\ln(1-\gamma/I)},
\end{equation} 
for illustration, see figure \ref{fig:MS_approach}f-j.
 
\begin{figure}[h]
    \centering
    \includegraphics[scale=1]{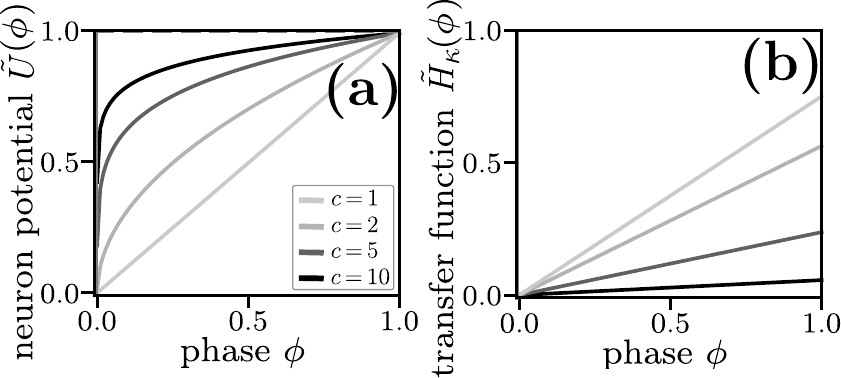}
    \caption{
    \textbf{Nonlinear neuron with linear transfer function for multiplicative coupling.} (a) While the rise function $\tilde{U}^{\mathrm{MS, }c}(\phi)$ (see eq. \eqref{eq:c_U}) itself is nonlinear, its respective transfer function for  multiplicative pulse coupling,   $\tilde{H}_{\kappa}(\phi)$, is linear and allows for easy analytical treatment. Shown for different values of $c$, for fixed coupling strength $\kappa=0.25$.
\label{fig:MSc}
}
\end{figure}

Note that the Mirollo-Strogatz neuron potential $\tilde{U}^{\mathrm{MS, }b}_\epsilon (\phi) = \frac{1}{b}\mathrm{ln}(1 + (e^b-1)\phi)$ as discussed in the last section 
does not lead to an affine transfer function anymore, if we apply multiplicative coupling:
\begin{equation}\label{eq:MS_H_prop}
\tilde{H}_{\kappa}^{\mathrm{MS, }b}(\phi) = \frac{1}{e^b-1}\left[(1+(e^b-1)\phi)^{(1-\kappa)}-1\right].
\end{equation}
However, we propose an alternative neuron potential,
\begin{equation}\label{eq:c_U}
\tilde{U}^{\mathrm{MS, }c}(\phi) \coloneqq \phi^{1/c}, \qquad c>0,
\end{equation}
which indeed leads to an affine (and even linear) transfer function
\begin{equation}
\tilde{H}^{\mathrm{MS, }c}_{\kappa}(\phi) = (1-\kappa)^c \phi
\end{equation}
for multiplicative coupling, see figure \ref{fig:MSc}a,b.

\section{equivalence of multiplicative and additive Coupling}

While with additive coupling (eq. \eqref{eq:update_rule}, \eqref{eq:update_rule_MS})
the state variable $x_i$ of a spike-receiving neuron $i$ is shifted by a constant value  $-\epsilon_{ij}$, with multiplicative coupling (eq. \eqref{eq:update_rule_prop}, \eqref{eq:MS_H}) the state variable $\tilde{x}_i$ experiences a relative shift $-\tilde{x}_i \kappa_{ij}$.
However, both interaction mechanism can be transferred into each other by simultaneously modifying the rise function accordingly. 
For sake of simplicity we focus only on a single neuron $i$ at a time and drop index $i$ throughout, and assume that all pulses arriving at neuron $i$ have the same strength:  $\epsilon_{ij} = \epsilon_i \equiv   \epsilon $ or  $\kappa_{ij} = \kappa_i \equiv  \kappa$. Below, we generalize to arbitrary coupling strength, and thus also sparse networks, see section \ref{subsec:inhomo}.

Again writing $U(\phi)$  and $\tilde{U}(\phi)$ for rise functions for additive and multiplicative coupling, respectively, we demand that the corresponding transfer functions
\begin{equation}
    {H}_{\epsilon}(\phi) = {U}^{-1}\left[{U}(\phi)-\epsilon \right]
    \label{eq:defH}
\end{equation} 
and 
\begin{equation}
 \tilde{H}_{\kappa}(\phi) = \tilde{U}^{-1}\left[\tilde{U}(\phi)  (1-\kappa)\right]   \label{eq:defHtilde}
\end{equation}
 are equal, i.e.
\begin{align}\label{eq:Hequivalent}
 {H}_{\epsilon}(\phi) & \equiv \tilde{H}_{\kappa}(\phi) \\
\iff  {U}^{-1}\left[{U}(\phi)-\epsilon \right] & = \tilde{U}^{-1}\left[\tilde{U}(\phi)  (1 - \kappa)\right],  \label{eq:equivalence_condition}
\end{align}
for all $\phi$.

\textbf{{Main claim.}}
As a main statement of this article we have the following:
Given a rise function $U(\phi)$ for additive inhibitory pulse coupling with coupling strength $\epsilon \geq 0$, exactly the same spiking dynamics is realized by multiplicative inhibitory pulse coupling with coupling strength $\kappa \in [0,1)$ and rise function 
\begin{equation}
  \tilde{U} (\phi ) = (1-\kappa)^{\left(\frac{1-U(\phi)}{\epsilon}\right)}.
  \label{eq:pot_trafo_abs_rel}
\end{equation}

To demonstrate this claim, we take several steps.
First, we calculate the inverse $\tilde{U}^{-1}$ of the rise function  $\tilde{U}$ (eq. \eqref{eq:pot_trafo_abs_rel}) with regard to variable $\phi$, expressed in terms of the inverse function $U^{-1}$ and $\tilde{U}$ itself,
\begin{equation}\label{eq:pot_trafo_abs_rel_inverse_of_U}
\tilde{U}^{-1} = U^{-1}\left( 1 - \epsilon \mathrm{log}_{1-\kappa}(\tilde{U})\right),
\end{equation}
 where $\mathrm{log}_{1-\kappa}$ denotes the logarithm with base $1-\kappa$.
We then use equations
 \eqref{eq:pot_trafo_abs_rel} and \eqref{eq:pot_trafo_abs_rel_inverse_of_U} to compute 
\begin{align}\label{eq:end_proof}
   \tilde{U}^{-1}&\left[\tilde{U}(\phi)  (1 - \kappa)\right]  =  \nonumber U^{-1}\left( 1 - \epsilon \mathrm{log}_{1-\kappa}(\tilde{U}(\phi) (1 - \kappa))\right) \\ \nonumber
   & =   U^{-1}\left( 1 - \epsilon \mathrm{log}_{1-\kappa}\left( (1-\kappa)^{\left(\frac{1-U(\phi)}{\epsilon}+1\right)}  \right)\right)\\\nonumber
   & =   U^{-1}\left( 1 - \epsilon  {\left(\frac{1-U(\phi)}{\epsilon}+1\right)}  \right) \\
   &= {U}^{-1}\left[{U}(\phi)-\epsilon \right].
\end{align}
Hence,  indeed the transfer functions, and thus the spiking dynamics, are identical, ${H}_{\epsilon}(\phi) = \tilde{H}_{\kappa}(\phi)$, see equation \eqref{eq:Hequivalent}.

By solving equation \eqref{eq:pot_trafo_abs_rel} for $\tilde{U} (\phi ) $, we get the inverse transformation. Given a multiplicative-coupling rise function $\tilde{U} (\phi )$, the same system can be described in terms of additive pulse coupling with rise function
\begin{equation}
U(\phi) = 1 - \epsilon \mathrm{log}_{1-\kappa}(\tilde{U}(\phi)).
\label{eq:pot_trafo_abs_rel_inverse}
\end{equation}

Equations \eqref{eq:pot_trafo_abs_rel} and \eqref{eq:pot_trafo_abs_rel_inverse} can be interpreted as families of equivalent rise functions.
For a given rise function $U(\phi)$ for additive coupling and a given additive coupling strength $\epsilon$, the coupling strength $\kappa$ of the transformed dynamics can be chosen freely, when $\tilde{U}$ is chosen accordingly, so that $\kappa$  effectively parametrizes the possible equivalent multiplicative-coupling rise functions $\tilde{U}$ .
Inversely, possible transformations to additive coupling are captured by different choices of the new coupling strength $\epsilon$, given a fixed original multiplicative-coupling strength $\kappa$.
For heterogeneous coupling, we refer to section \ref{subsec:inhomo}.

While transformations   \eqref{eq:pot_trafo_abs_rel}, \eqref{eq:pot_trafo_abs_rel_inverse} are given in terms of the rise functions $U(\phi)$ and $\tilde{U}(\phi)$, 
they are   easily transferred to the actual neuron dynamics by letting $   x(t) = U(\phi(t))$, $   \tilde{x}(t) = \tilde{U}(\phi(t))$:
\begin{equation}
  \tilde{x} (t) = (1-\kappa)^{\left(\frac{1-x(t)}{\epsilon}\right)}
  \label{eq:x_trafo},
\end{equation}
and
\begin{equation}
x(t) = 1 - \epsilon \mathrm{log}_{1-\kappa}(\tilde{x} (t))
\label{eq:x_trafo_inv},
\end{equation}
where we assume $x(T) = \tilde{x}(T)=x^\mathrm{thr} = 1$. 
We point out that the period length $T$ and hence also the phase velocity $d\phi/dt = \omega = 1/T$ is necessarily the same for equivalent rise functions.

\subsection{Additive to multiplicative coupling}

\begin{figure*}[!]
    \centering
    \includegraphics[scale=2]{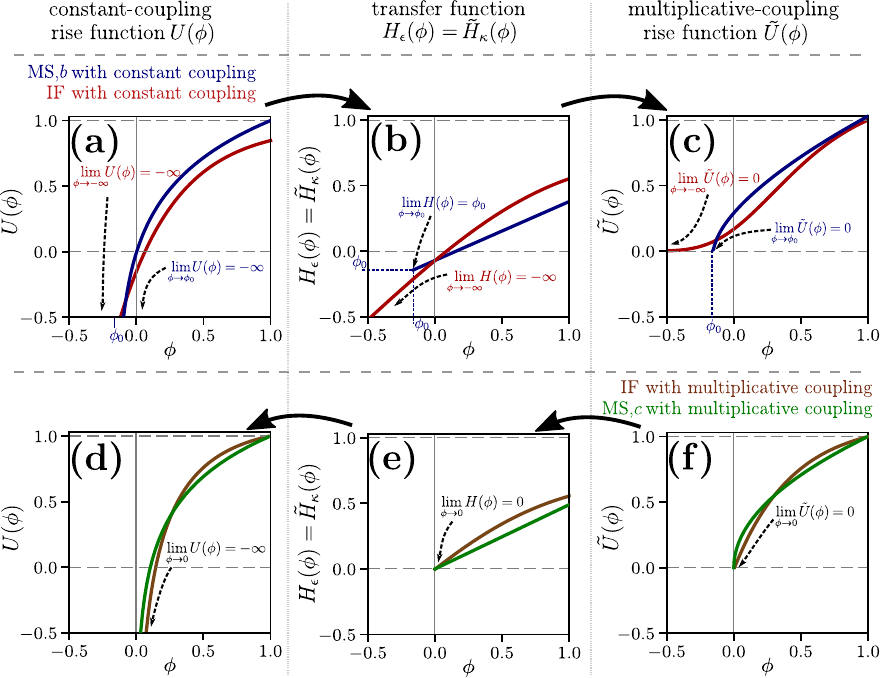}
    \caption{
    \textbf{Transformation between additive and multiplicative pulse coupling.}
    (a,d)  Additive-coupling rise function $U(\phi)$.
    (b,e)  Transfer function $H_\epsilon(\phi) = \tilde{H}_\kappa(\phi)$.
    (c,f) Multiplicative-coupling rise function   $\tilde{U}(\phi)$.
(red) Leaky integrate-and-fire neuron model with additive coupling  (eq. \eqref{eq:IF_H} with $I=1$, $\gamma=0.9$) , $\epsilon=0.2$, $\kappa=0.3$.   
(blue) Mirollo-Strogatz neuron potential with additive coupling (eq. \eqref{eq:MS_U}, $b=2$) , $\epsilon=0.4$, $\kappa=0.4$.    
(brown): Leaky integrate-and-fire neuron model with multiplicative coupling (eq. \eqref{eq:IF_U_mult} with $I=1$, $\gamma=0.9$), $\epsilon=0.2$, $\kappa=0.2$.  (green): Neuron potential $\tilde{U}^{\mathrm{MS, }c} $ with linear transfer function for multiplicative coupling (eq. \eqref{eq:c_U_2} for $c=2$),  with $\epsilon = 0.2$, $\kappa = 0.3$.
\label{fig:trafo_main}
}
\end{figure*}

First we consider the transformation from additive to multiplicative coupling via equation \eqref{eq:pot_trafo_abs_rel}, as illustrated by figure \ref{fig:trafo_main}a-c  for two standard neuron models. Note that for the transformed neuron dynamics, $\tilde{U}(\phi)$, we have $\tilde{U}(0)>0$.
Indeed, from equation \eqref{eq:pot_trafo_abs_rel}   we find 
\begin{equation}\label{eq:U_R}
\tilde{U}_R = (1-\kappa)^{\frac{1 - U_R}{\epsilon},},
\end{equation}
where we write $\tilde{U}_R= \tilde{U}(0)$ and $U_R = U(0)$
 for the reset potentials of $\tilde{U}(\phi)$ and $U(\phi)$, respectively.
 According to equation \eqref{eq:U_R} it is not possible to normalize both to $U(0)=0$ and $\tilde{U}(0) = 0$ at the same time.
 In fact, it is a necessary condition for the equivalence between a multiplicative-coupling rise function $\tilde{U}$ and an additive-coupling rise function $U$ that $\tilde{U}(\phi_0) = 0$ and ${U}(\phi_0) = - \infty$
for the same phase $\phi_0$, or in the limit $\phi \to - \infty$. Illustratively speaking, $\tilde{U}(\phi_0) = 0$ and ${U}(\phi_0) = - \infty$ define the phase $\phi_0$ at which the effect of multiplicative and additive coupling, respectively, vanishes, which must be the same for equivalent dynamics.

If we require $U(0)=0$, as is usually done for additive-coupling rise functions, the reset value $\tilde{U}_R = \tilde{U}(0) = (1-\kappa)^{1/\epsilon} >0$ is always larger than the minimally allowed potential $ \tilde{U}(\phi) = 0$.
From a practical perspective, the equivalent multiplicative coupling uses a baseline potential $\tilde{U}_G = 0$ lower than the reset potential $\tilde{U}_R>0$[ which can be considered a natural generalization of the standard multiplicative coupling as introduced in the last section.]
In this case, $U_R = 0, \tilde{U}_R>0$, we can also rewrite the transformation
\eqref{eq:pot_trafo_abs_rel} 
  in terms of the reset voltage $\tilde{U}_R$ of the multiplicative coupling,
\begin{equation}\label{eq:trafo_UR}
\tilde{U}(\phi) = \tilde{U}_R^{(1-U(\phi))}.
\end{equation}
While in equation  \eqref{eq:pot_trafo_abs_rel}    the family of equivalent multiplicative-coupling rise functions is parametrized in terms   of the coupling strengths $\kappa$ of the new coupling, equation  \eqref{eq:trafo_UR} 
parametrizes the possible transformations in terms of $\tilde{U}_R$, from which the new coupling strength  is found as $\kappa = 1 - (\tilde{U}_R)^\epsilon$.

For example, consider the standard leaky integrate-and-fire potential (eq. \eqref{eq:IF_U})
  \begin{equation}\label{eq:IF_U_2}
 U^\mathrm{IF, add.}(\phi) = \frac{I}{\gamma}\left(1- e^{-\gamma\phi T}\right)  =  \frac{I }{\gamma}\left(1- \left(1-\frac{\gamma}{I }\right)^{\phi}\right),
 \end{equation} where $T = - \gamma^{-1} \textrm{ln}(1-{\gamma}/{I})$,
with an additive-coupling transfer function 
  \begin{equation}\label{eq:IF_H_2}
H^\mathrm{IF, add.}_{\epsilon}(\phi)   = \frac{\ln\left((1-\gamma/I)^{\phi} + \epsilon \gamma/I\right)}{\ln(1-\gamma/I)}.
 \end{equation}
As before, we use a tilde to differentiate between rise functions for multiplicative or additive pulse coupling; the superset \enquote{$\mathrm{add.}$} denotes that the original model (here integrate-and-fire, \enquote{$\mathrm{IF}$}) was formulated for additive coupling. 
Transforming via equation \eqref{eq:pot_trafo_abs_rel}, we get an equivalent rise function  
\begin{align}\label{eq:U_tilde_IF_const}
\tilde{U}^\mathrm{IF, add.} (\phi) & = (1-\kappa)^{\frac{1}{\epsilon}
\left(
1- \frac{I}{\gamma}\left(
1-(1-\frac{\gamma}{I}
\right)^\phi
\right)} \\
& = \tilde{U}_R^{
\left(
1- \frac{I}{\gamma}\left(
1-(1-\frac{\gamma}{I}
\right)^\phi
\right)}
\end{align}
for multiplicative coupling.

As another common example for additive coupling we consider the rise function 
\begin{equation}
U^{\mathrm{MS, }b\mathrm{, add.}}(\phi) = \frac{1}{b}\mathrm{ln} \left(
1 + \left(e^b - 1 \right) \phi
 \right), \qquad b>0,
\end{equation} (see eq. \eqref{eq:MS_U})
which 
diverges for a finite phase   $\phi_0 =  1/(1-e^b) \in  (-\infty, 0]\binoppenalty=\maxdimen
\relpenalty=\maxdimen
$ and
corresponds to an additive-coupling transfer function
\begin{equation}
H^{\mathrm{MS, }b\mathrm{, add.}}_\epsilon (\phi) = \frac{e^{-b \epsilon} - 1}{e^b - 1} + e^{-b\epsilon} \phi,
\end{equation}
(see eq. \eqref{eq:MS_H})
and an equivalent multiplicative coupling rise function
\begin{align}
\tilde{U}^{\mathrm{MS, }b\mathrm{, add.}} (\phi) &= \left(
1 + \left(
e^b - 1)\phi
\right)
\right)^{\frac{\ln{(1-\kappa)}}{b \epsilon}} \\ \nonumber
&= \left(
1 + \left(
e^b - 1)\phi
\right)
\right)^{b^{-1}\ln{\tilde{U}_R}}
\end{align}

We point out that for practical applications, the offset reset voltage, $\tilde{U}_R>0$, might be avoided by modifying the transformed rise function $\tilde{U}(\phi)$ in an interval $\phi \in [0, \delta]$ arbitrary close to the reset phase $\phi=0$, such that $\tilde{U}(0) = 0$ and $\tilde{U}(\phi)$ is still strictly monotonous and continuous. While doing so might slightly change the collective dynamics during transients, for stable orbits in which neurons do not reach non-positive phases via inhibition,  $\delta$ can be chosen such that the dynamics are completely equivalent to the original additive coupling.

\subsection{Multiplicative to additive coupling}

Figure \ref{fig:trafo_main}d-f illustrates the transfer of multiplicative  to additive coupling via equation \eqref{eq:pot_trafo_abs_rel_inverse} in the case of $\tilde{U}(0) = \tilde{U}_R \equiv 0\binoppenalty=\maxdimen
\relpenalty=\maxdimen
$. This  represents the most natural formulation for multiplicative coupling (as described in the last section, also see figure \ref{fig:MSc}), where
the reset potential $\tilde{U}_R$ is equal to the ground potential $\tilde{U}_G = 0$.
In this case, the effect of the coupling vanishes right at the reset phase $\phi =  0 $. 
Hence, no negative phases are necessary, $\phi \in [0,1]$. 
However, multiplicative coupling with $ \tilde{U}_R = 0$  requires  a divergence of the additive-coupling rise function $U(\phi)$ at $\phi=0$,
 $\lim_{\phi \to 0^+} U(\phi) = -\infty$, forbidding normalization to $U_R = U(0) \equiv 0$, or any other finite value.

Note that the phase $\phi_1$, above which the neuron potential for additive coupling becomes positive, $U(\phi_1) = 0$, is connected to the coupling strength $\epsilon$ via $(1-\kappa)^{1/\epsilon} = \tilde{U}(\phi_1)$, so that it can be chosen freely within $(0, 1)$ by adjusting $\epsilon$ accordingly, in principle arbitrary close to $0$. Instead of $\epsilon$, we can also use $\phi_1$ or $\tilde{U}(\phi_1)$ to parametrize the family of equivalent additive-coupling transfer functions:
\begin{equation} \label{eq:phi_1}
    U(\phi) = 1- \mathrm{log}_{\tilde{U}(\phi_1)}\tilde{U}(\phi) 
    = 1 - \frac{\mathrm{ln} \, \tilde{U}(\phi) }{\mathrm{ln}\, \tilde{U}(\phi_1) }
\end{equation}

Consider, for example, multiplicative coupling with a standard leaky integrate-and-fire rise function 
\begin{equation}
\tilde{U}^\mathrm{IF , mult.}(\phi) = 
    \frac{I }{\gamma}\left(1- \left(1-\frac{\gamma}{I }\right)^{\phi}\right),
\end{equation}
and a corresponding transfer function

\begin{equation}
\tilde{H}_{\kappa}^\mathrm{IF, mult.}(\phi) =  \frac{\ln\left(  
\kappa + (1-\kappa)    (1-\gamma/I)^\phi 
\right)}{\ln(1-\gamma/I)},
\end{equation}  
where we use the superscript \enquote{$\mathrm{mult.}$} to point out that the original rise function is implemented for multiplicative coupling.
 Via equation \eqref{eq:pot_trafo_abs_rel_inverse} we find an equivalent additive-coupling rise function
\begin{align}
U^\mathrm{IF, mult.}(\phi) & = 1 - \epsilon \log_{1-\kappa} \left(\frac  {1-(1-  \gamma/I)^ \phi } { \gamma/I} \right ) \\
&= 1 -  \log_{\tilde{U}(\phi_1)} \left(\frac  {1-(1-  \gamma/I)^ \phi } { \gamma/I} \right),\nonumber
\end{align} see figure \ref{fig:trafo_main}g-i.

As second example we give 
\begin{equation}\label{eq:c_U_2}
\tilde{U}^{\mathrm{MS, }c\mathrm{, mult.}}(\phi) = \phi^{1/c}, \qquad c>0,
\end{equation}
with a linear multiplicative-coupling transfer function
\begin{equation}
\tilde{H}^{\mathrm{MS, }c\mathrm{, mult.}}_{\kappa}(\phi)  = (1-\kappa)^c \phi
\end{equation}
and equivalent additive coupling rise function 
\begin{align}
U^{\mathrm{MS, }c\mathrm{, mult.}} (\phi) & = 1 - \frac{\epsilon}{c} \log_{1-\kappa} \left(\phi \right) \\
& =  1 - \frac{\ln{(\phi)}}{\ln{(\phi_1)}},\nonumber
\end{align}
which is the Mirollo-Strogatz neuron potential $U^{\mathrm{MS, }b}$ (\mbox{eq. \eqref{eq:MS_U})} shifted such that its divergence occurs at  $\phi_0 = 0$.

Additive-coupling rise functions as created by the transformation \eqref{eq:pot_trafo_abs_rel_inverse} might be  sufficiently approximated by  rise function avoiding divergent behavior as $\phi \to 0^+$,   for $\phi \to 0^+$, allowing normalization to $U(0)=0$. For a practical example, see   section \ref{sec:kwta}.

\subsection{Inhomogeneous coupling strengths}\label{subsec:inhomo}
In the last sections we demonstrated a general approach to transfer multiplicative pulse-coupling to additive pulse-coupling and vice-versa.
By setting $\epsilon_{ij} \equiv  \epsilon$ and $\kappa_{ij}  \equiv \kappa$, we implicitly   assumed that the coupling strengths $\epsilon_{ij}$ do not depend on the neuron $j$ which is sending the pulse, but only on the receiving neuron $i$ (whose index $i$ we suppressed throughout).
However, the equivalence between multiplicative and additive coupling holds also for inhomogenuous coupling strengths and coupling  topologies defined on networks.
In order to have equivalent dynamics for variable original coupling strengths, not only the rise functions, but also the coupling strengths itself are transformed via a mapping that is uniquely determined by defining two values of $\epsilon$ and $\kappa$ as equivalent.
Consider equation \eqref{eq:pot_trafo_abs_rel} again:
\begin{equation}
  \tilde{U} (\phi ) = (1-\kappa)^{\left(\frac{1-U(\phi)}{\epsilon}\right)}.
  \label{eq:pot_trafo_abs_rel_1}
\end{equation}
The corresponding transfer functions for additive and multiplicative coupling are equal not only for the specific choice of $\epsilon$, $\kappa$ used in the transformation \eqref{eq:pot_trafo_abs_rel_1} itself, but also for any combination of additive and multiplicative coupling strengths $\epsilon'$ and $\kappa'$ that satisfies
\begin{equation}\label{eq:kappaprime}
    \kappa' = 1 - (1 - \kappa)^{\epsilon' / \epsilon},
\end{equation}
as can be verified analogously to equations \eqref{eq:defH}-\eqref{eq:end_proof}.
Hence, for each neuron $i$, a single choice of two equivalent coupling strengths defines both the transformed rise functions as well as the mapping between arbitrary equivalent coupling strengths $\epsilon_{ij}$, $\kappa_{ij}$.

For transforming standard additive coupling with $U_R = U(0)=0$ to multiplicative coupling,
the mapping between equivalent coupling strengths may also be defined in terms of the reset voltage $\tilde{U}_R = \tilde{U}(0)$ of the new rise function, not requiring an explicit choice of two equivalent coupling strengths. The new multiplicative-coupling rise function is then given in terms of 
\begin{equation}\label{eq:trafo_UR_1}
\tilde{U}(\phi) = \tilde{U}_R^{(1-U(\phi))},
\end{equation}
see equation \eqref{eq:trafo_UR},
while the mapping between equivalent coupling strengths $\epsilon_{ij}$, $\kappa_{ij}$
is given by 
\begin{equation}
    \kappa_{ij} = 1 - \tilde{U}_R^{(\epsilon_{ij})}.
\end{equation}

Similarly, if the possible transformations of multiplicative coupling with $\tilde{U}_R=0$ to additive coupling are parametrized via the phase $\phi_1$ for which $U(\phi_1)=0$,
\begin{equation} \label{eq:phi_1_1}
    U(\phi) = 1- \mathrm{log}_{\tilde{U}(\phi_1)}\tilde{U}(\phi) 
    = 1 - \frac{\mathrm{ln} \, \tilde{U}(\phi) }{\mathrm{ln}\, \tilde{U}(\phi_1) },
\end{equation} see equation \eqref{eq:phi_1}), 
the transformation between equivalent coupling strengths $\kappa_{ij}$, $\epsilon_{ij}$ is defined via 
\begin{equation}
   {\epsilon_{ij}} = \frac{\log{(1-\kappa_{ij})}}{ \log{(\tilde{U}(\phi_1))}}.
\end{equation}

Interaction topologies defined on a directed  graph can be either implemented explicitly in terms of an adjacency matrix $A_{ij}$ that defines which neurons $i$ receive a pulse from another neuron $j$ or, implicitly, by setting $\epsilon_{ij} =  0$ or equivalently $\kappa_{ij} = 0$ for neurons that are not connected in the specific direction.
For an example of delayed heterogeneous coupling on a directed network, see the next section.

\section{Example Applications}
 
 \subsection{Network computing}\label{sec:kwta}
 
To give a practical example for a transformation from multiplicative to additive pulse-coupling, we consider an implementation of a $k$-winners-take-all ($k$-WTA) computation based on an  network of $N$ inhibitorily coupled oscillatory neurons, as proposed in \cite{Neves2020ReconfigurableComputation}. The free dynamic of neuron  $i \in \{1, \dots, N\}$ with state variable $\tilde{x}_i$
is defined by differential equation 
\begin{equation}\label{eq:kwta_eq}
    \dot{\tilde{x}}_i = \left( I -  \gamma \tilde{x}\right) \xi_i(t),
\end{equation} with constants $I>\gamma$ and external inputs $\xi_i(t): \mathbb{R} \to \mathbb{R}^+$,
and a reset rule, which sets ${\tilde{x}}_i(t_i^+)\coloneqq 0$ for each time $t_i$ where ${\tilde{x}}_i(t_i) = {\tilde{x}}^\mathrm{thr} \equiv 1$. Assuming that $\xi_i(t)$ changes much slower than the inputs $x_i(t)$, we get a  free dynamic
\begin{equation}
\tilde {x}_i(t) = \frac{I}{\gamma  \xi_i(t)}(1-e^{-\gamma t}) \qquad \mathrm{for} \qquad 0<t\leq T,
\end{equation}
with free period length $T = - (\xi_i(t) \gamma)^{-1} \textrm{ln}(1-{\gamma}/{I})$.
The interaction between different neurons is given in terms of all-to-all inhibitory multiplicative pulse-coupling 
\begin{equation}
\tilde{x}_i(t_j^+) = (1 - \kappa)\,\tilde{x}_i(t_j)  ,
\end{equation} (cf. eq. \eqref{eq:update_rule_prop})
with global coupling strength $\kappa$.
 
\begin{figure*}[!]
    \centering
    \includegraphics[scale=1.95]{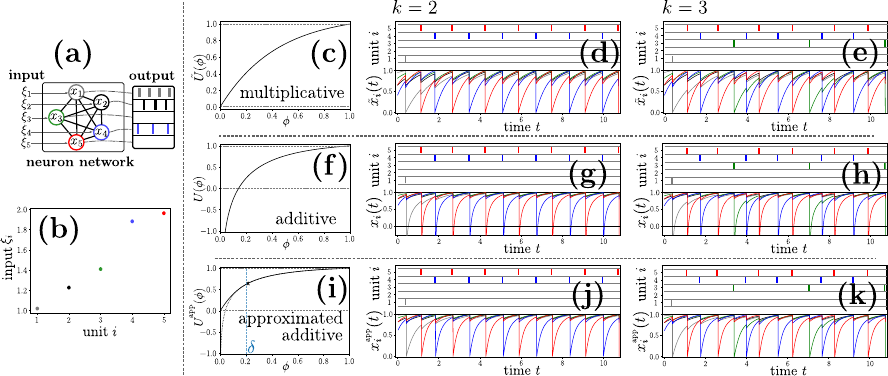}
    \caption{\textbf{Reconfigurable $k$-WTA computation via symmetrical inhibitory coupling for multiplicative and additive coupling.}
    (a) All-to-all network of $N=5$ oscillatory neurons with time-dependent external inputs $\xi_i(t)$ setting the intrinsic frequencies of the individual neurons. (b): Choice of $\xi_i$ for the illustrated dynamics. 
    (c,f,i) Rise functions for multiplicative, additive, and approximated additive coupling with $U^\mathrm{app}(0)=0$. In (i), The phase $\delta$ below which $U^\mathrm{app}(\phi)$ (solid line) deviates from $U(\phi)$ (dotted line) is denoted by a blue line. The minimum phase $\phi$ that is reached via inhibition is marked by a cross.
    (d,g,j) Equivalent  dynamics with $k=2$ spiking neurons for multiplicative, additive, and approximated additive coupling. $\kappa = 2$ for multiplicative and $\epsilon = 0.1$ for additive coupling.
    (e,h,k) Equivalents dynamics with $k=3$ spiking neurons for multiplicative, additive, and approximated additive coupling. $\kappa' = 1.9$ for multiplicative and $\epsilon' \approx 0.089$ for additive coupling.
\label{fig:ex_kwta}
}
\end{figure*}

Now, for performing $k$-WTA computations, the   functions $\xi_i(t)$, which set the intrinsic frequencies \hbox{$\omega_i= 1/T_i \propto \xi_i(t)$} of the individual neurons, are interpreted as the time-dependent input of the network,  while the stream of reset events (i.e. the sent pulses) define the output space, see figure \ref{fig:ex_kwta}a.
Upon varying $\xi_i(t)$, the system after a typically short transient converges to a periodic orbit, in which only the  $k \leq N$ neurons with the largest intrinsic frequencies $\omega_i$  spike, thereby revealing the subset of the $k$ largest input signals $\xi_i$.
The number of \enquote{winners} $k$ is selected by adjusting the global coupling strength $\kappa$ accordingly, allowing for easy re-configurability.
The underlying mechanism is summarized as follows: Upon receiving a pulse from a neuron $i$, due to the multiplicative inhibitory coupling all other neurons $j\neq i$ loose a certain share $\kappa$ of their voltage, pushing their voltages closer together on an absolute scale. Depending on the inhibition strength, this allows faster neurons to overtake slower ones repeatedly, potentially keeping the latter from spiking altogether.
Figure \ref{fig:ex_kwta}d,e illustrates typical dynamics with $k=2$ and $k=3$, for a specific choice of input signals $\xi_i(t)=\xi_i$ (fig. \ref{fig:ex_kwta}b) in the original multiplicative-coupling formulation, with rise function
 \begin{equation} 
\tilde{U_i}(\phi_i)  = \frac{I }{\gamma}\left(1- \left(1-\frac{\gamma}{I }\right)^{\phi_i}\right),
 \end{equation} (see fig. \ref{fig:ex_kwta}c) and transfer function 
\begin{equation} 
\tilde{H}_{\kappa, i}(\phi_i) =   \frac{\ln\left(  
\kappa + (1-\kappa)    \left((1-\frac{I}{\gamma}\right)^{\phi_i }
\right)}{\ln\left(1-\frac{I}{\gamma} \right)}.
\end{equation} 

The equivalent dynamics (fig. \ref{fig:ex_kwta}g,h)   for additive coupling are implemented via a free neuron dynamic given by 
\begin{equation}\label{eq:kwta_trafo_x}
     x_i(t) = 1 - \epsilon \frac{\log{\left(\frac{I}{\gamma}-( \frac{I}{\gamma}-(1-\kappa)^{1/\epsilon} e^{(t\gamma)} \right)}}
     {\log{(1-\kappa)}},
\end{equation}
which corresponds to an additive-coupling rise function
\begin{equation}\label{eq:kwta_trafo}
     U_i(\phi_i) = 1 - \epsilon \frac{\frac{I }{\gamma}\left(1- \left(1-\frac{\gamma}{I }\right)^{\phi_i}\right)}
     {\log{(1-\kappa)}},
\end{equation} see figure   \ref{fig:ex_kwta}f. 

Note that  neither $\tilde{U_i}(\phi_i)$ nor $U_i(\phi_i)$   depend on $\xi_i(t)$ itself because for the specific  free dynamics as given by equation \eqref{eq:kwta_eq} the shape of the rise function itself does not change with $\xi_i(t)$, but the  intrinsic neuron frequency $\omega_i = 1/T_i$).
For the illustrated transformation, we required that multiplicative coupling strength $\kappa=0.21$ (for the first orbit, with $k=2$) is equivalent to additive coupling strength $\epsilon=0.1$,  so that  $\kappa'=0.19$ (for the second orbit, with $k=3$) is equivalent to \hbox{ $\epsilon' = \epsilon \log{(1-\kappa')}/\log{(1-\kappa)} \approx 0.089$}, (cf. eq. \eqref{eq:kappaprime}).

Figure \ref{fig:ex_kwta}i-k  illustrates how additive coupling with $U(0)= U_R = -\infty$ can be effectively approximated by a modified additional-coupling rise function $U^\mathrm{app}(\phi)$ which satisfies $U^\mathrm{app}(0)=0$ for a  more practical implementation.
Here, we require that $U^\mathrm{app}(\phi) \equiv U(\phi)$ for phases $\phi \geq \delta \equiv 0.2$ and define  $U^\mathrm{app}(\phi)$ for $\phi < \delta$ in terms of cubic extrapolation  which satisfies $U^\mathrm{app}(0)=0$.
For our specific choice of $\delta=0.2$, the illustrated spiking dynamics are in fact fully equivalent to both the original multiplicative and the exact additive transformation, because the minimal phase reached upon inhibition is larger than $\delta=0.2$ (see fig. \ref{fig:ex_kwta}i).

 \subsection{Topology-induced synchronization}
 
 As a second example, we consider a network with delayed, heterogeneous coupling,  implementing a phenomenon initially addressed in \cite{timme2006topology} which relates topological features of directed interaction graphs with the degree of synchronization in the resulting collective dynamics.
Though originally described for continuously coupled Kuramoto oscillators, we here reproduce the  effect for leaky-integrate and fire-dynamics with delayed additive inhibitory pulse coupling and then exemplarily transfer the dynamics to multiplicative coupling.
 \begin{figure*}[t]
    \centering
    \includegraphics[scale=2.0]{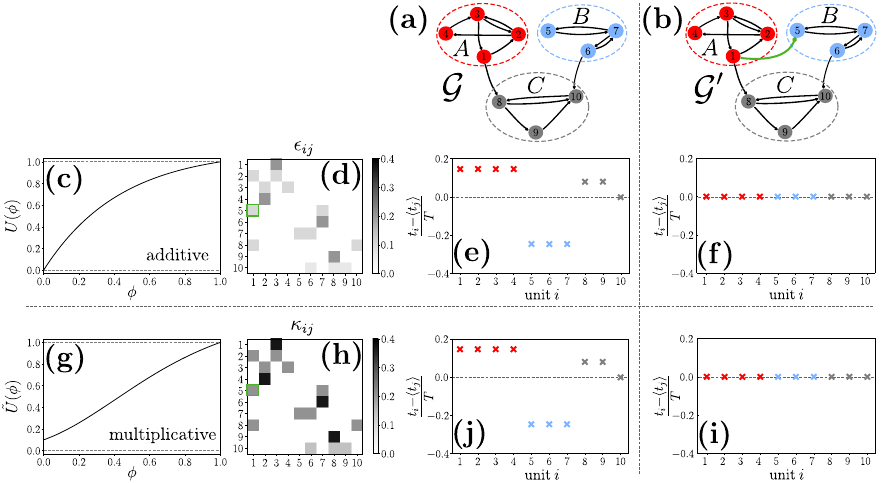}
    \caption{\textbf{Topology-induced phase synchronization via inhibitory coupling with delay, for multiplicative and additive coupling.} 
    (a) Network of $N=10$ identical integrate-and-fire neurons, with interaction topology defined by graph $\mathcal{G}$. Strongly connected components $A$ and $B$ affect $C$ independently from each other. (b) After addition of a directed link from node $1$ to $5$, $A$ affects both $B$ and $C$ and acts as a single \enquote{source} of the new interaction graph $\mathcal{G}'$. (c,g) Equivalent rise functions $U(\phi)$ and $\tilde{U}(\phi)$ for additive and multiplicative pulse coupling. For additive coupling, we use standard integrate-and-fire dynamics as described by eq. \eqref{eq:IF_U}, with $I=1$, $\gamma=0.9$. 
    (d,h) Equivalent coupling strengths $\epsilon_{ij}$ and $\kappa_{ij}$ for additive and multiplicative coupling, respectively, as given by eq. \eqref{eq:kappaeps}, for the interaction graph $\mathcal{G}$. The matrix element corresponding to the added link is marked by a blue square. For the delay  between all connected units we take $\tau = 1$.
    (e,j) Phase lag of the individual units $i$ relative to the period length $T$  (periodic) collective dynamic after a transient. Strongly connected components $A$ and $B$ act as individual sources of the directed interaction graph $\mathcal{G}$, not allowing for global phase synchronization.
    (f,g) Adding a single link makes $A$ the only source in the new interaction  graph $\mathcal{G}$, leading to full phase synchronization.
\label{fig:FIG_ex_kuramoto}
}
\end{figure*}

 Consider a network of $N$ identical oscillatory neurons with a leaky integrate-and-fire dynamic
\begin{equation}\label{eq:kuramoto_eq}
    \dot{{x}}_i =   I -   \gamma {{x_i}},
\end{equation} with $I=1$, $\gamma=0.9$,
and a reset rule which sets \hbox{${{x}}_i(t_i^+) = 0$} for each time $t_i$ where ${{x}}_i(t_i) = {{x}}^\mathrm{thr} \equiv 1$, leading to a free time evolution
\begin{equation}
x_i(t) = \frac{I}{\gamma}(1-e^{-\gamma t}) \qquad \mathrm{for} \qquad 0<t\leq T,
\end{equation}
with  period length $T = - \gamma^{-1} \textrm{ln}(1-{\gamma}/{I})$.
The interaction between different units is given in terms of delayed inhibitory additive pulse-coupling with 
\begin{equation} \label{eq:kuramoto_coupling}
 {x}_i(t_j^+ + \tau) =  {x}_i(t_j + \tau) - \epsilon_{ij},
\end{equation} (cf. eq. \eqref{eq:update_rule}), where $\tau>0$.
The coupling strengths $\epsilon_{ij}$ are given in terms of the adjacency matrix $A$ of an directed graph with matrix elements $A_{ij} \in \{0,1\}$  as
\begin{equation}
     \epsilon_{ij}= \epsilon A_{ij}/g_i ,
\end{equation}
 where $g_i = \sum_j A_{ij}$ is the in-degree of node $i$ and $\epsilon>0$ is a global constant.
Figure \ref{fig:FIG_ex_kuramoto}a,b  depicts two interaction graphs, $\mathcal{G}$ and $\mathcal{G}'$, which are  identical except for an additional directed link between neurons $1$ and $5$ in the second graph. 

Figure \ref{fig:FIG_ex_kuramoto}e,f shows for both network topologies the relative time lag between reset events of the neurons after a periodic orbit is reached, from the same random initial condition.
While the first interaction topology leads to only partial phase synchronization (fig.  \ref{fig:FIG_ex_kuramoto}e), adding a single link, $A_{5,1}\coloneqq 1$,  leads to full synchronization with no phase lag at all (fig.  \ref{fig:FIG_ex_kuramoto}f).
This is directly connected to the strongly-connected components of the graphs $\mathcal{G}$ and $\mathcal{G}'$, referred to as $A$, $B$, and $C$. While in the first interaction topology, $A$ and $B$ independently affect $C$, with no feedback from $C$ back to $A$ or $C$ and no interaction at all between $A$ and $B$, in the second topology, component $A$ additionally affects $B$.
While for $\mathcal{G}$, $A$ and $B$ synchronize independently to different phases depending on the initial conditions  and do not allow for $C$ to synchronize to a single phase, for $\mathcal{G}'$,
$A$ acts as a single \enquote{source} of the system, which synchronizes autonomously and forces its phase on the rest of the network via directed connections  to all other nodes.

In terms of the phase formalism, for the original additive coupling (eq. \eqref{eq:kuramoto_coupling}), we have rise functions
 \begin{equation}\label{eq:IF_U_}
 U_i(\phi) = \frac{I}{\gamma}\left(1- e^{-\gamma\phi_i T}\right)  =  \frac{I }{\gamma}\left(1- \left(1-\frac{\gamma}{I }\right)^{\phi_i }\right),
 \end{equation}
and  transfer functions
   \begin{equation}\label{eq:IF_H_}
H_{i, \epsilon_{ij}}^{\mathrm{IF}}({\phi_i })   = \frac{\ln\left((1-\gamma/I)^{\phi_i } + \epsilon_{ij} \gamma/I\right)}{\ln(1-\gamma/I)}.
 \end{equation}
 Note that the neurons itself are identical in their free dynamics, and only differ in their coupling strengths, and hence their transfer functions, see figure \ref{fig:FIG_ex_kuramoto}c.
 Although it is not strictly necessary, it is reasonable to maintain identical free dynamics also  when using multiplicative coupling, resulting in different coupling strengths $\kappa_{ij}$ for each pulse-receiving neuron $i$.
 Hence, we set the reset voltage $\tilde{U}_R = \tilde{U}(0)>0$ to be the same for each neuron, $\tilde{U}_R \equiv 0.1$ and use it to define the transformation  
 to multiplicative coupling (see eq. \eqref{eq:trafo_UR}), leading to free neuron dynamics
\begin{equation}
    \tilde{x}_i(t) = \tilde{U}_R^{\left(1-\frac{I}{\gamma}(1-e^{-\gamma t})\right)} 
\end{equation}
 or, equivalently, rise functions
\begin{equation}
   \tilde{U}_i  (\phi_i)    = \tilde{U}_R^{
\left(
1- \frac{I}{\gamma}\left(
1-(1-\frac{\gamma}{I}
\right)^{\phi_i}
\right)},
\end{equation} see figure 
\ref{fig:FIG_ex_kuramoto}g
with corresponding multiplicative coupling strengths
\begin{equation}\label{eq:kappaeps}
    \kappa_{ij} = 1 - \tilde{U}_R^{(\epsilon_{ij})} = 1-   \tilde{U}_R^{\epsilon A_{ij}/g_i
    },
\end{equation}
see figure \ref{fig:FIG_ex_kuramoto}d,h.
Starting from equivalent initial conditions, the resulting spiking dynamic and hence also the degree of phase locking (see fig. \ref{fig:FIG_ex_kuramoto}j,i) is exactly the same as for the original additive-coupling formulation.

 \section{Conclusion}

 Basic standard models of spiking neural networks such as networks of leaky integrate-and-fire neurons exhibit additive interactions where the postsynaptic response of a neuron is parametrized in terms of a coupling strength. Other and more advanced models feature multiplicative coupling where the state change of the postsynaptic neuron depends on the state that neuron is in at the time of pulse reception.

Here we have demonstrated that under certain conditions, additive and multiplicative coupling may be viewed as two mathematically equivalent options for modeling. In particular, the phase representation originally introduced by Mirollo and Strogatz for additive coupling \cite{Mirollo1990Synchronization} is readily modified to equally characterize dynamics of pulse-coupled systems with multiplicative coupling. To explicate the equivalence most clearly, we analyze a simple class of systems with instantaneous
(delta-coupled) postsynaptic responses and homogeneous inhibitory coupling.
We find in particular that inhibitory multiplicative coupling can be transferred to additive coupling and vice versa by simultaneously modifying the neuron potential, that is, the free neuron dynamics, so that the resulting spiking dynamics are identical.
The coupling parameter of the new coupling may be chosen freely, if the neuron's rise function is selected accordingly.
We discuss some peculiarities of the transformed neuron models in detail, such as non-zero reset voltages, and suggest approximations to avoid these.
The equivalence between additive and multiplicative coupling holds also for inhomogeneous coupling as well as networked systems with intricate interaction topology, if all coupling strengths are transformed by the same mapping (for every receiving neuron).

To illustrate the range of applicability, we simulate two different collective phenomena both for models of additive and multiplicative coupling.
The first example supports $k$-winner-takes-all computations via symmetrically all-to-all coupled neuron networks; the second topology-dependent synchronization for inhibitory pulse coupling with delays.
The findings explicate that indeed exactly the same collective dynamics can be generated by models with either type of coupling.
Also, we exemplify that slight
manipulations of the transformed neuron potentials can allow more effective realization, often without changing the resulting collective dynamics at all.

These results not only advance the theoretical foundations for modeling spiking neural networks but may also help transfering knowledge about systems with additive to systems with multiplicative coupling and vice versa. Such complementary results may in particular ease software or hardware implementations of spiking neural networks for different purposes.

\bibliography{lit_auto}

\end{document}